\documentclass[a4paper,11pt,amsmath,amssymb]{article}
\usepackage{pos}

\def\beq{\begin{equation}} \def\eeq{\end{equation}}
\def\beqn{\begin{eqnarray}} \def\eeqn{\end{eqnarray}}

\newcommand\aem{\alpha} 
 
\def\beq{\begin{equation}} 
\def\eeq{\end{equation}} 
\def\beqn{\begin{eqnarray}} 
\def\eeqn{\end{eqnarray}} 
 
\def\to{\rightarrow}

\pdfoutput=1

\newcommand{\valencia}{Instituto de F\'{\i}sica Corpuscular, Universitat de Val\`{e}ncia -- Consejo Superior de Investigaciones Cient\'{\i}ficas, Parc Cient\'{\i}fic, E-46980 Paterna, Valencia, Spain.}
\newcommand{\culiacan}{Facultad  de  Ciencias  F\'isico-Matem\'aticas,  Universidad  Aut\'onoma  de  Sinaloa,  Ciudad  Universitaria, CP 80000 Culiac\'an, M\'exico.}
\newcommand{\berlin}{Deutsches Elektronen-Synchrotron DESY, Platanenallee 6, 15738 Zeuthen, Germany.}

\begin{document}

\title{Elucidating the internal structure of hadrons through direct photon production}
\author[a]{David F. Renter\'ia-Estrada}
\author[a]{Roger J. Hern\'andez-Pinto,}
\author[b,c]{German F. R. Sborlini}
\affiliation[a]{\culiacan}
\affiliation[b]{\berlin}
\affiliation[c]{\valencia}

\emailAdd{davidrenteria.fcfm@uas.edu.mx}
\emailAdd{roger@uas.edu.mx}
\emailAdd{german.sborlini@desy.de}

%

\abstract{The accurate description of the internal structure of hadrons is a very challenging task. In order to compare the predictions with the highly-accurate experimental data, it is necessary to control any possible source of theoretical uncertainties. Thus, we can use the information extracted from final state measurement to constrain our knowledge about the internal structure of hadrons. In this work, we describe how direct photon production can be exploited to unveil details about the partonic distributions inside protons. Also, we explain how to describe QCD-QED corrections to hadron plus photon production at colliders, focusing on the accurate reconstruction of the partonic momentum fractions from experimentally accessible observables.}

\FullConference{%
  *** The European Physical Society Conference on High Energy Physics (EPS-HEP2021), ***\\
  *** 26-30 July 2021 ***\\
  *** Online conference, jointly organized by Universität Hamburg and the research center DESY ***
}

\setcounter{page}{1}
\maketitle

\section{Introduction}
\label{sec:introduction}
One of the best strategies for studying the internal structure of the non-fundamental particles is based on the parton model. It relies on factorization properties and parton distribution functions (PDFs) that allow the characterization of the hadronic structure \cite{Collins:1989gx}. The PDFs are obtained from the experiments, for example from the Large Hadron Collider (LHC), through advanced model fitting methods. However, this methodology is not sufficient to explain the total spin of the proton. 
Using the currently available data, only 30\% of the total spin can be explained from the quarks contributions~\cite{EuropeanMuon:1989yki,HERMES:2006jyl,COMPASS:2006mhr}. Furthermore, data extracted from deep inelastic scattering (DIS) experiments 
are not sufficient to place restrictions on the polarized distributions of quarks and gluons \cite{deFlorian:2008mr, deFlorian:2009vb}.

There are different methodologies to characterize the internal structure of hadrons. 
A method that gives direct access to the dynamics of the partons and obtain more information about the proton's spin, consists of detecting the photons created by the internal interaction of the partons within the hadron: these are the so-called \emph{hard} photons. The proton-proton collision creates a dense medium with very high temperatures: as a consequence, any particle that couples to QCD partons will interact with this medium. Due to the non-perturbative nature of this interactions, it is usually preferable to detect particles that weakly interacts with this medium: this is why we will focus on the identification of hard photons.

In this work, we analyze the production of one direct photon accompanied by a hadron, 
including leading-order (LO) corrections in QED and 
next-to-leading order (NLO) corrections in QCD. Specifically, we analyze the effects of QED corrections on the distribution of the cross-section, for the energies of the PHENIX (200 GeV) and LHC (13 TeV) experiments \cite{Renteria-Estrada:2021rqp}. This complements the phenomenological study presented in Ref. \cite{deFlorian:2010vy} and constitutes a step forward towards a better understanding of the process and an updated description of the phenomenology of hadron-photon production at colliders.


\section{Cross-section calculation}
\label{sec:calculation}
In the parton model, thanks to the factorization theorem~\cite{Collins:1989gx}, the cross-section can be described by the convolution between parton distribution functions, fragmentation functions (FFs) and the partonic cross-section. The non-perturbative effects associated with the low energy interactions are included in the PDFs and FFs, while the partonic cross-section is calculated from a perturbative framework, i.e.
\beqn
d\sigma_{H_1 \, H_2 \to h \, \gamma}^{\rm DIR} &=& \sum_{a_1 a_2 a_3} \int dx_1 dx_2 dz \, f^{(H_1)}_{a_1}(x_1,\mu_I) f^{(H_2)}_{a_2}(x_2,\mu_I) \, d^{(h)}_{a_3}(z,\mu_F) d\hat\sigma^{\rm DIR}_{a_1\,a_2 \to a_3 \, \gamma}  \, .
\label{eq:Direct}
\eeqn
The PDF is represented by $f_a^{(H)}(x,\mu_I)$, describing the probability density of finding a parton $a$ inside a hadron $H$, with momentum fraction $x$ and initial energy scale $\mu_I$. Similarly, the FF, $d_{a_3}^{h}(z,\mu_F)$, represents the probability density of generating one hadron $h$ with momentum fraction $z$ at the energy scale $\mu_F$.

To estimate the cross-section we considered two possible scenarios: \textbf{i)} the \emph{direct} contribution (i.e. the photon is generated directly from the partonic interaction), and \textbf{ii)} the \emph{resolved} contribution (the photon comes from the fragmentation of a hadron). Due to the quantum nature of the process, it is not possible to identify the true origin of the detected photon. To mitigate this problem we implement the smooth cone isolation algorithm~\cite{Frixione:1998jh}. This is a good method that give the efficiently separate the direct from the resolved component. Due to the high correlation between the production of hard photons and isolated photons, we implemented the procedure described in Ref.\cite{Frixione:1998jh} to identify the hard photons. The smooth cone isolation algorithm allows for soft gluons throughout space, and ensures an IR-safe cross-section. Finally, under these assumptions, Eq. (\ref{eq:Direct}) can be written as
\beqn
d\sigma_{H_1 \, H_2 \to h \, \gamma} &=& \sum_{a_1 a_2 a_3} \int dx_1 dx_2 dz \, f^{(H_1)}_{a_1}(x_1,\mu_I) f^{(H_2)}_{a_2}(x_2,\mu_I) \, d^{(h)}_{a_3}(z,\mu_F) d\hat\sigma^{\rm ISO}_{a_1\,a_2 \to a_3 \, \gamma}  \, ,
\label{eq:Isolated}
\eeqn
where $d\hat\sigma^{\rm ISO}_{a_1\,a_2 \to a_3 \, \gamma}$ is the partonic cross-section that contains the smooth cone isolation prescription. There are two partonic channels that contribute at LO, 
\beq
q \bar q \to \gamma g \, , \quad q g \to \gamma q \, ,
\label{eq:PartonicChannelsLO}
\eeq
while at NLO we must also consider:
\beq
q \bar q \to \gamma g g \, , \quad q g \to \gamma g q \, , \quad g g  \to \gamma q \bar q \, , \quad q \bar q \to \gamma Q \bar Q \, , \quad q Q \to \gamma q Q ~.
\label{eq:PartonicChannelsNLO}
\eeq
To implement NLO QCD corrections we rely on the FKS subtraction algorithm  \cite{Frixione:1995ms}, which divides the real emission phase-space into different regions containing non-overlapping singularities and cancels them with the local IR counter-terms. Renormalization was carried out within the $\overline{\rm MS}$ scheme. 
Also, to estimate LO QED contribution, we add
\beqn
 d\hat\sigma^{\rm ISO, QED}_{a_1\,a_2 \to a_3 \, \gamma} &=& \frac{\aem^2}{4\pi^2}\, \int d{\rm PS}^{2\to 2} \,  \frac{|{\cal M}^{(0)}_{QED}|^2(x_1 K_1, x_2 K_2, K_3/z, K_4)}{2 \hat s} \, {\cal S}_2 \,  ,
\label{eq:xsISOLATEDQED}
\eeqn 
to Eq. (\ref{eq:Isolated}). The new partonic channels involved in the QED corrections are $q \gamma \to \gamma q$ and $q \bar q \to \gamma \gamma$, where the photon PDF appears.



\section{Results}
\label{sec:results}
\begin{figure}[t!]
    \centering
    \includegraphics[width=0.47\textwidth]{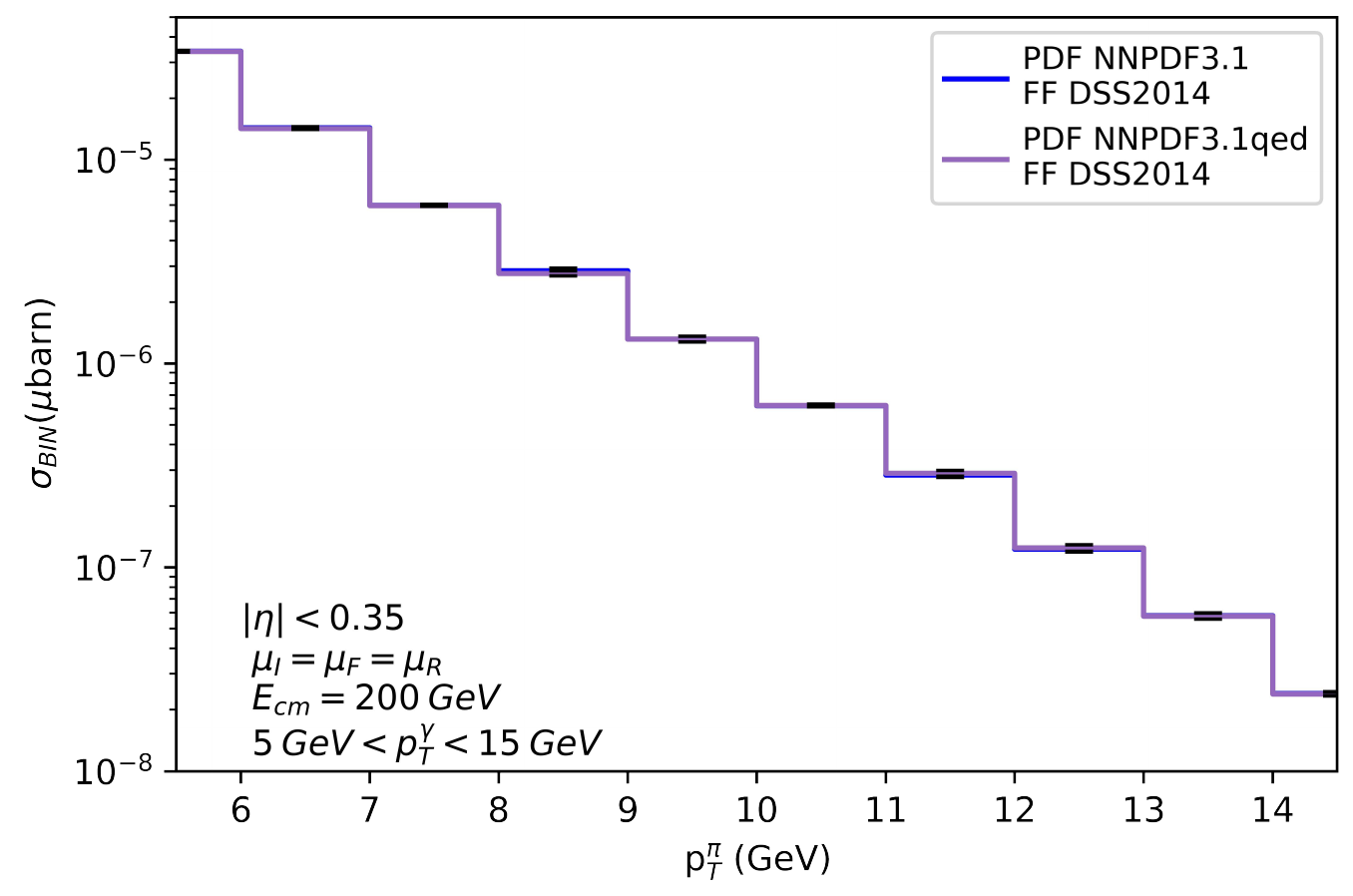} \ \includegraphics[width=0.47\textwidth]{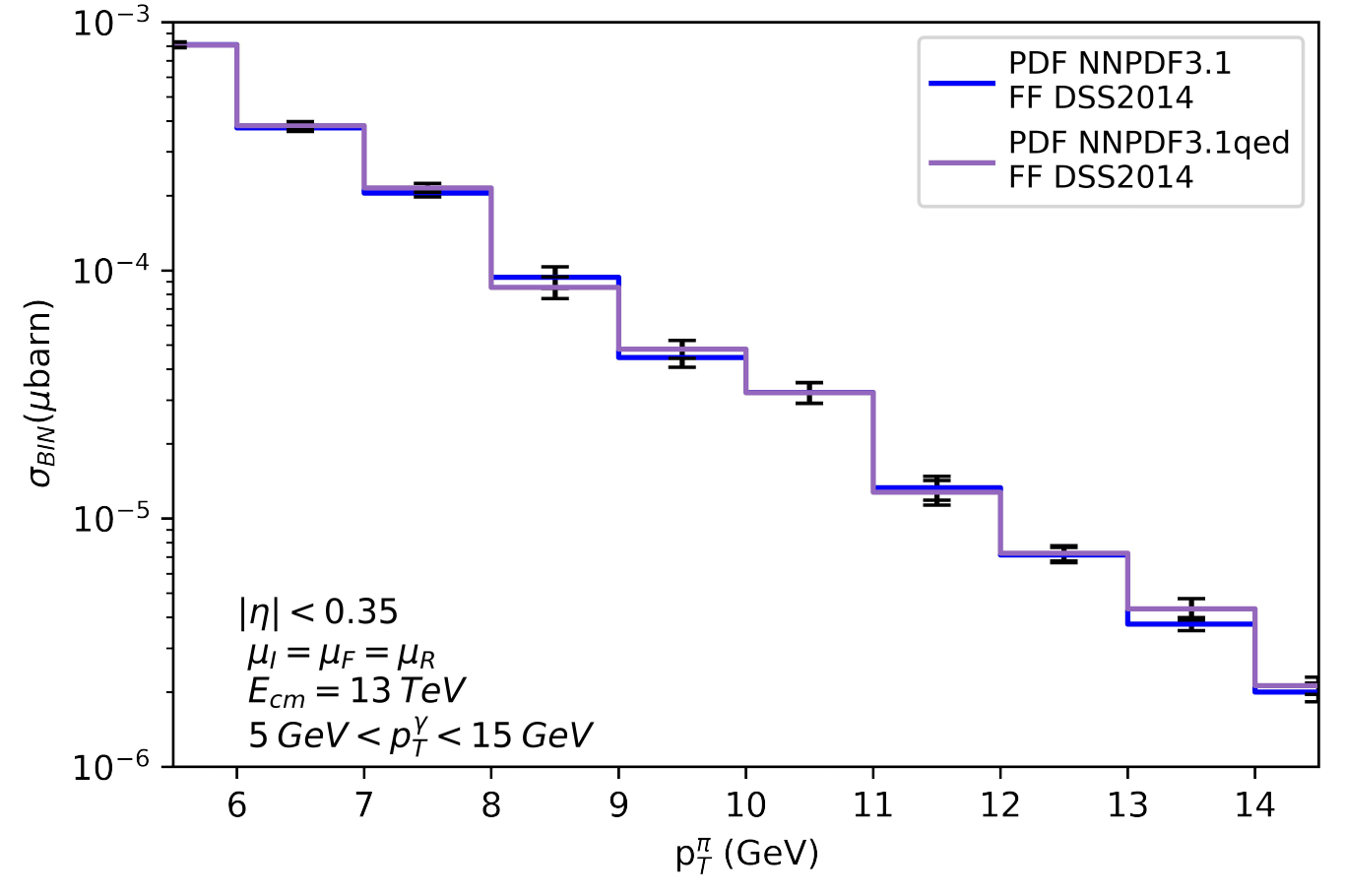}
    \caption{Comparison of the combined NLO QCD + LO QED (purple line) versus the pure NLO QCD corrections to the $p_T^{\pi}$ distribution of $p+p \to \pi^++\gamma$. In the left, we consider PHENIX kinematics, whilst we switched to LHC energies in the right plot.}
    \label{fig:Figura1}
\end{figure}

We focus on the process $p\,+\,p\, \to\, h\, + \, \gamma$, specifically looking for a charged pion in the final state. We implement a Monte-Carlo code with the following default kinematic cuts: 
\begin{enumerate}
    \item The rapidities of the photon and pion are restricted to $ |\eta| \leq 0.35 $.
    \item The transverse momentum of the photon is restricted to $5 \, {\rm GeV} \leq p_T^{\gamma} \leq 15 \, {\rm GeV}$. 
    \item The transverse momenta of the pion must be greater than 2 GeV. 
    \item We do not consider restrictions on the azimuthal angles of the photon and the pion. We constrain $\Delta \phi=|\phi^\pi - \phi^\gamma|\geq 2$ (to include those events where the photon and pion are produced close to a back-to-back configuration). 
\end{enumerate}
These cuts simulate PHENIX experiment at RHIC. We have explored center-of-mass energies of $E_{CM} = 200 {\rm GeV}$ (PHENIX) and $E_{CM} = 13 {\rm TeV}$ (LHC). Furthermore, we used more modern sets of PDFs and FFs than those in the work of Ref.~\cite{deFlorian:2010vy}. Then, we explored the impact of NLO QCD and LO QED corrections as a function of the center-of-mass energy. We implemented the \texttt{LHAPDF} interface~\cite{Buckley:2014ana} to select different PDF sets. For the default configuration, the NLO QCD correction is fixed using the PDF set \texttt{NNPDF3.1}. To analyse the effects of the combined LO QED + NLO QCD corrections, we use \texttt{NNPDF3.1luxQEDNLO} \cite{Manohar:2017eqh,Buonocore:2020nai}~. Finally, we update the FF routines to implement the recent set \texttt{DSS2014}~\cite {deFlorian:2014xna}. In Fig. \ref{fig:Figura1} we compare the process for different center-of-mass energies. As expected from the behaviour of the electromagnetic coupling, QED corrections are more relevant with increasing of center-of -mass energy, as well as in the high $p_T$ region. 

\section{Conclusions}
\label{sec:conclusions}
In this work, we have studied the phenomenology of photon-hadron production. We included LO QED + NLO QCD corrections and performed a careful comparison with the QCD only contributions. We used the \texttt{NNPDF3.1luxQEDNLO} set, centering the analysis on the $p_T^\pi$ spectrum. We found small but non-negligible corrections: ${\cal O}(2\, \%)$ for PHENIX and ${\cal O}(8\, \%)$ for LHC center-of-mass energies. As already discussed in other works \cite{deFlorian:2010vy,Renteria-Estrada:2021rqp}, these results suggest the hadron plus photon might be an important process to impose tighter constraints on PDFs. In particular, using novel computational techniques (such as machine learning and artificial intelligence), we could expect to have a better understanding of the partonic kinematics by using the photon as a probe to explore the dynamics of the collisions~\cite{DAVIDinpreparation}.

\section*{Acknowledgements}
We would like to thank P. Zurita for reading this manuscript and fruitful discussions for future works. This research was partially supported by COST Action CA16201 (PARTICLEFACE). The work of D. F. R.-E. and R. J. H.-P. is supported by CONACyT through the Project No. A1- S-33202 (Ciencia Basica). Besides, R. J. H.-P. is also funded by Ciencia de Frontera 2021-2042 and Sistema Nacional de Investigadores from CONACyT.

\providecommand{\href}[2]{#2}\begingroup\raggedright\endgroup

\end{document}